\documentclass[]{llncs}

\pdfoutput=1

\usepackage{graphicx}
\usepackage{float}
\usepackage{color}
\usepackage[T1]{fontenc}
\usepackage{textcomp}
\usepackage{gensymb}
\usepackage{siunitx}
\usepackage{array}
\usepackage{tabularx}
\usepackage{colortbl}
\usepackage{hhline}
\usepackage{multirow}
\usepackage{balance}
\usepackage{subfigure}
\usepackage{listings}
\usepackage{booktabs}
\usepackage{textcomp}
\usepackage{xspace}

\newcommand{\fakeparagraph}[1]{\vspace{0.1em}\noindent\textbf{#1}.}

\newcommand{\eg}{\emph{e.g.},\xspace}

\begin{document}

\title{Delay-Tolerant Networking for Long-Term Animal Tracking}

 \author{Philipp Sommer \and Branislav Kusy \and Philip Valencia \and Ross Dungavell \and Raja Jurdak}
 
\institute{CSIRO, Autonomous Systems, Brisbane, QLD, Australia\\\email{firstname.lastname@csiro.au}}

\date{\vspace{-5ex}}

\maketitle

\begin{abstract}
Enabling Internet connectivity for mobile objects that do not have a permanent home or regular movements is a challenge due to their varying energy budget, intermittent wireless connectivity, and inaccessibility. We present a hardware and software framework that offers robust data collection, adaptive execution of sensing tasks, and flexible remote reconfiguration of devices deployed on nomadic mobile objects such as animals. The framework addresses the overall complexity through a multi-tier architecture with low tier devices operating on a tight energy harvesting budget and high tier cloud services offering seamless delay-tolerant presentation of data to end users. Based on our multi-year experience of applying this framework to animal tracking and monitoring applications, we present the main challenges that we have encountered, the design of software building blocks that address these challenges, and examples of the data we collected on flying foxes.
\end{abstract}
\section{Introduction}

Tracking wildlife at continental scale has received high research interest for decades due its significance for ecological conservation and disease spread monitoring. Ecologists are interested in understanding the movement and condition of animals across different landscapes and environmental conditions, as this understanding can lead to better conservation and management decisions~\cite{Kays2015}. The significance of wildlife tracking is further amplified for species that can carry virulent and potentially deadly diseases, as the movement of animals correlates highly with the likelihood that the disease spreads across the landscape~\cite{Jurdak15}. For instance, fruit bats are known carriers of viruses such as Nipah, Ebola, and Hendra. Australian federal and state governments have set up a national monitoring program to understand where these highly mobile animals travel for better understanding of Hendra virus dynamics, through a combination of individual animal tracking tags and manual surveys of roosting sites.
 
Effectively tracking wildlife involves temporal, spatial, and functional aspects. On the temporal dimension, ecologists need individual tracking of animals that is long-term, high frequency, and delay-tolerant. Long-term operation is required because the animals move in nomadic patterns that vary over time scales of weeks to months.  High frequency position and condition sampling are important for fine-grained understanding of animal movement patterns and behaviour. Delay-tolerant operation ensures that the data from tagged animals with prolonged disconnections from the communication infrastructure is reported once connectivity returns. Spatially, tracked animals can move large distances over short time periods. For instance, flying foxes can typically move tens of kilometres over a night, and have been reported to move up to 600 kilometres in one night. The communication infrastructure for retrieving data from tracked animals must therefore be spatially spread across large geographical areas and must provide seamless and efficient download capabilities. Finally, at an operational level, wildlife tracking is clearly a long-term activity where many of the research questions evolve with inflow of new data, which necessitates highly reconfigurable and re-taskable systems. 

Recent work in wildlife tracking, however, falls short on addressing the above requirements. Most tracking projects collect either short-term frequent data or long-term sparse data, subject to very limited battery energy that is constrained by the weight of tracking nodes. Delay-tolerance has been partially addressed in some efforts by logging data continuously and offloading it when connections resume~\cite{Lindgren2008}. For tracking systems over large spatial scales, however, base stations that serve as the data collection points typically act independently, without coordinating with other base stations to check whether data from the tracking node had been previously downloaded. This creates potential for redundancies in data download as well as bandwidth and energy inefficiencies in the system. Finally, currently available systems do provide a degree of manual remote reconfiguration of sensor sampling schedules, such as GPS, but they do not support re-tasking or fully reprogramming nodes remotely. This limits the versatility of tracking studies once the nodes are deployed. 

This article addresses the above challenges by proposing an architecture for long-term delay-tolerant networking, consisting of three tiers: (1) mobile nodes, (2) gateways, and (3) cloud services. Our mobile nodes include solar panels, GPS, and many low-power sensors, in addition to algorithms for energy-based sensor scheduling and delay tolerant data collection. Our gateway nodes that are spatially dispersed across an area spanning more than 2000\,km (see Figure~\ref{fig:example}(a)) synchronise data availability through the cloud services tier to ensure that data is downloaded only once throughout a continental-scale deployment. The architecture supports, in addition to remote reconfiguration, full remote reprogramming of the nodes once they are in contact with a base station, providing maximum versatility for long-term tracking studies. We showcase the features of our architecture through the motivating application of tracking flying foxes across Australia, which has inspired its original design. 
\section{Related Work}

Technological advances in positioning systems have been adopted by scientists to facilitate the study of animal behaviour in their natural habitat. Starting from the 1960s, animals have been tagged with VHF radio transmitters, which allowed to determine the animal's location using triangulation from multiple receiver locations. However, this process is very labour intensive, often requiring biologists to walk through the animal's habitat, and thus potentially affecting the animal's behaviour.

With the launch of the Argos satellite system in 1978 it became possible to track platform transmitter terminals (PTT) by measuring Doppler shifts~\cite{seegar1996fifteen}, which allowed for global tracking applications with high spatio-temporal resolution. Beacon signals received by the Argos satellites are forwarded to ground-based processing stations and can then be accessed by Telnet.

The availability of the Global Positioning System (GPS) for civilian applications in the 1990s provided for the first time very accurate positioning for animal tracking devices. Driven by advances in consumer electronics, form factor and power consumption of GPS receivers have further decreased since then. Modern Argos transmitters have also been combined with an integrated GPS receiver to improve location accuracy, but still use satellites to relay positioning data. 

Within the last decade, wireless radio transceivers have been integrated into tracking devices to offload positioning samples to a base station located within the habitat~\cite{Zhang04} or employ cellular communication networks (GSM/GPRS/3G)~\cite{Anthony2012} to relay position and sensor information in near real-time. 
\section{Challenges}
Animal ethics considerations push limits of the form factor and weight of electronic devices attached to animals and deployment-of-scale requirements limit the unit cost of tracking devices. Consequently, tracking devices will have limited computation, communication, data storage, and energy capacity available on the device. Here, we distill the key technical challenges that we encountered in long-term animal tracking applications.

\fakeparagraph{Challenge \#1: Constrained and Variable Energy Budget} Due to stringent constraints for weight and form factor, mobile nodes need to rely on tiny batteries for energy storage. Most current work in animal tracking employs non-rechargeable batteries for high frequency sampling~\cite{Franzmitter,Dyo2010}. Energy harvesting allows to recharge the on-board batteries and thus allows for long-term operation. However, energy harvesting exposes new challenges with energy availability and sensor sampling as energy budgets are dynamic and unpredictable with an order of magnitude difference between energy harvested on sunny summer and cloudy winter days. Therefore, the software needs to gracefully adapt to a varying energy budget. To avoid missing critical data due to lack of energy, the nodes need to schedule sampling of several on-board sensing modalities in accordance with the current battery state of charge, daily energy budget, and predicted activity of the mobile object. 

\fakeparagraph{Challenge \#2: Intermittent Network Connectivity} Wildlife can roam vast areas each day, which are often remote and not within coverage of cellular communication networks. Satellite-based tracking systems provide global coverage, but have limited bandwidth. High subscription costs and weight/energy constraints restrict their use in some animal tracking applications. While traditional commercial devices~\cite{Franzmitter} based on VHF radio offer opportunistic wireless download through a portable receiver device, labor costs of such human-based data collection methods are prohibitive for large deployments. Animal tracking projects have mainly focused on either a single base station~\cite{Zhang04,Butler04}, which limits their spatial scalability, or collection of contact logs~\cite{Lindgren2008} without capturing heterogeneous sensor data, or multi-modal sensing without absolute position sampling~\cite{Dyo2010}. Instead, our software framework is designed to offer seamless communication with local data buffering to support situations when animals leave a known area for weeks and return to a different area, possibly hundreds of kilometers away.

\fakeparagraph{Challenge \#3: Lack of Physical Access} Due to the nature of the deployment scenarios involving animals, physical access to nodes might not be possible after the initial deployment, as it is often infeasible to capture a tagged animal again. Therefore, mobile nodes need to operate on a near-perpetual basis for long periods without physical human intervention. However, our experience has shown that it is often necessary to verify that nodes operate correctly and to refine the initially selected sensing parameters as more data gets available. Therefore, it is important that the software framework provides methods for remote debugging and task configuration that operate over the wireless channel and handle intermittent connectivity between gateways and mobile nodes. 

\fakeparagraph{Case Study} Our case study of flying fox tracking exemplifies this scenario. Flying foxes, also known as fruit bats or mega bats, congregate in large numbers in day roosts (bat camps), where placing a gateway node provides a great opportunity to download data from the tagged animals in the roost. During nightly foraging flying foxes can fly large distances and migrate to other roosts. While many animals come in proximity with a gateway placed at a known day roosting camp every few days, it might also take several weeks before the next contact with a gateway is made.
\section{Network Architecture}
\label{sec:architecture}

\begin{figure}[tbh]
\centering
\includegraphics[width=0.9\textwidth]{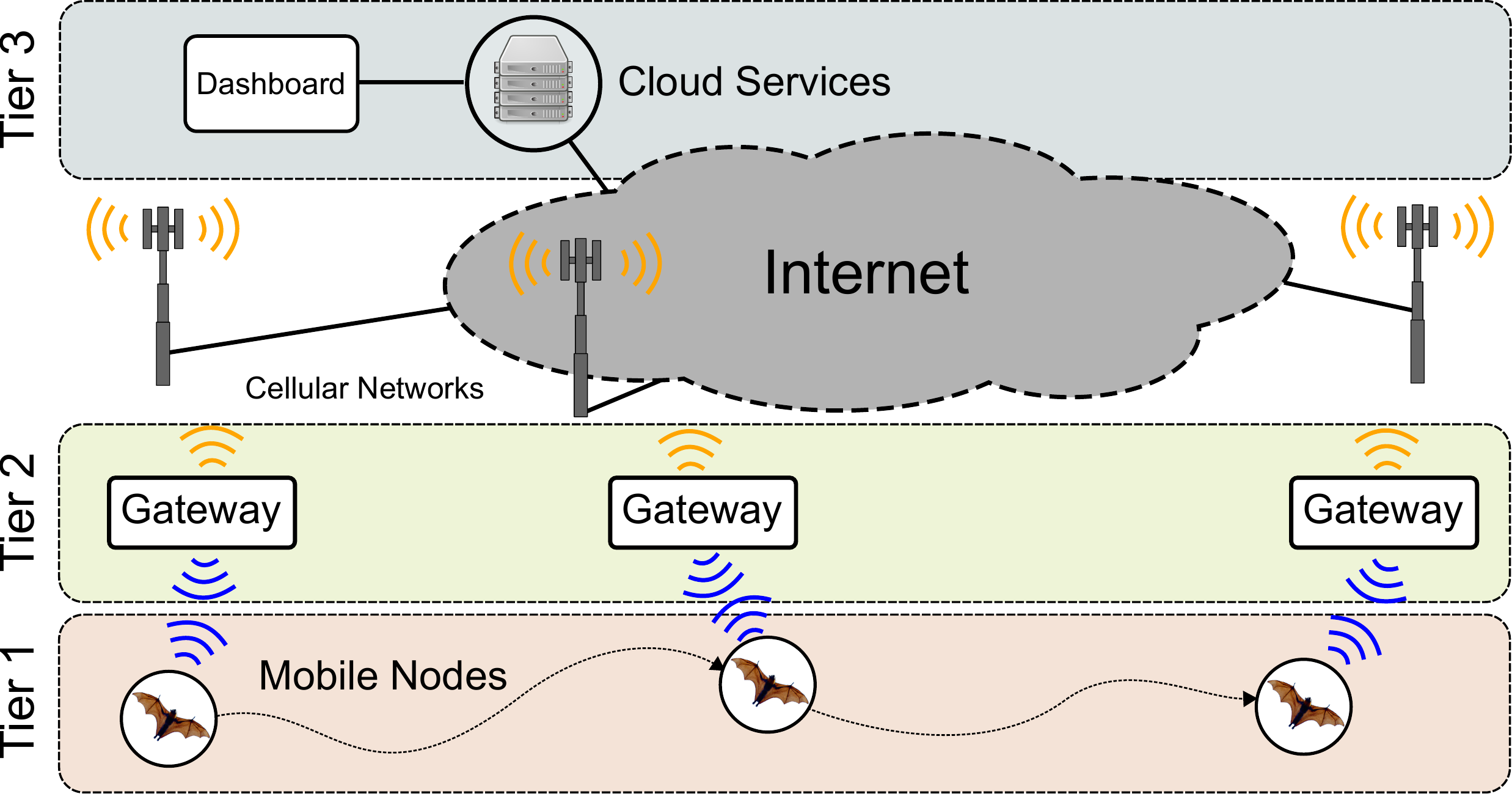}\vspace*{0.5cm}
\includegraphics[height=0.32\textwidth]{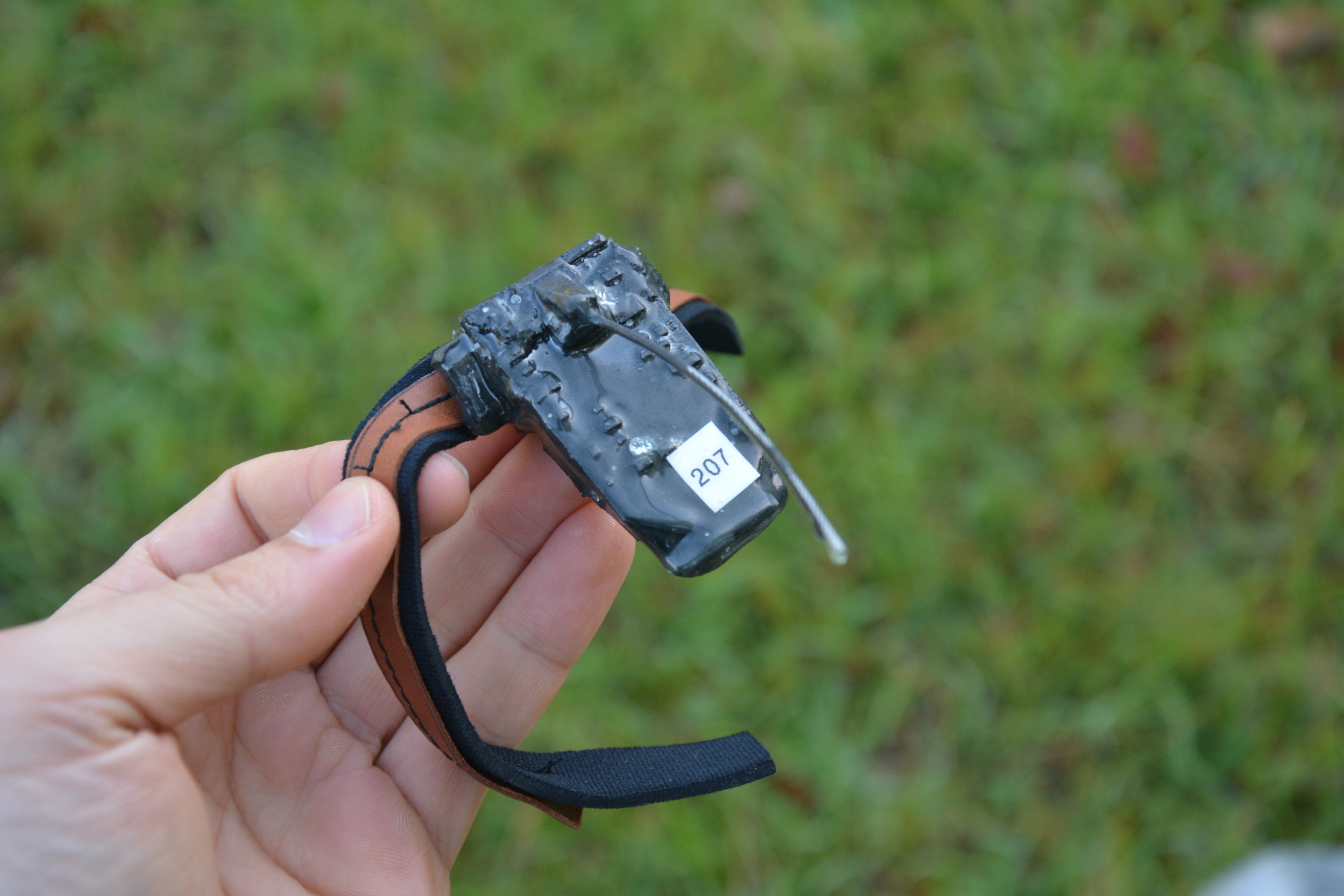}
\includegraphics[height=0.32\textwidth]{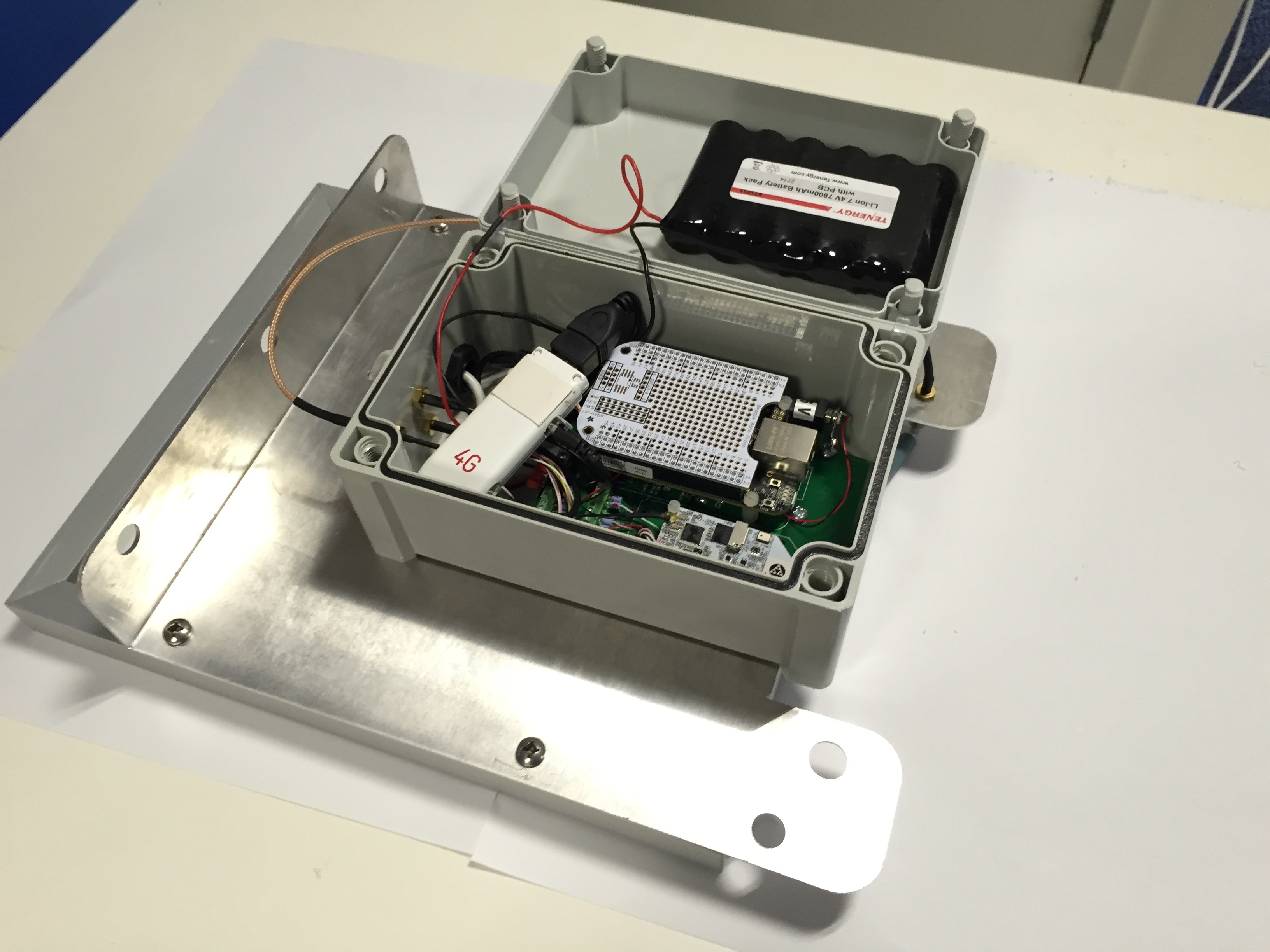}
\caption{Three-tier device architecture for the delay-tolerant animal tracking and monitoring (top). Mobile node integrated into a collar for flying foxes (bottom left), and a gateway node mounted at the back of a solar panel (bottom right).}
\label{fig:architecture}
\end{figure}

The main challenge that we need to tackle is the lack of wireless communication infrastructure in a majority of the animal's habitat. Consequently, we based our system architecture around a sparse network of gateways that communicate directly with mobile nodes to download the most recent data and use the Internet to connect to a cloud-based service to deliver sensor data and to synchronise the metadata among gateways (see Figure~\ref{fig:architecture}). 

\fakeparagraph{Mobile Nodes (Tier 1)} The mobile nodes are attached to the monitored animal using specially designed collars or halters. When designing the mobile node units, the form factor and weight restrictions heavily depend on the physical size of the tagged objects (\eg animal ethics regulations require collars to weigh less than 5\% of the animal's body weight, which is about 20\,g for the flying fox example). Mobile nodes feature sensor chips for multiple sensing modalities, persistent data storage and a short-range wireless transceiver~\cite{Jurdak2013}. We use the Texas Instruments CC430 system-on-chip together with Contiki OS~\cite{Dunkels2004}, which is optimised to prolong node lifetime when running from small batteries. Consequently, the energy budget, computational power, communication bandwidth and storage capabilities of such devices are highly constrained.

\fakeparagraph{Gateway Nodes (Tier 2)} The second tier consists of battery- and solar-powered gateway nodes, which are deployed at animal congregation areas and are equipped with a 900\,MHz short-range wireless radio to communicate with the mobile nodes. As the communication range of the mobile nodes is typically restricted to a few hundred meters, gateways are placed at strategic locations where animals tend to congregate (\eg flying foxes roosting camps, cattle drinking troughs) to maximise the opportunity of wireless connectivity with mobile nodes. Gateways are built around low-cost embedded platforms and can be connected to the Internet using cellular networks, or point-to-point WiFi links.

We implemented a framework for remote method invocations with low overhead using bi-directional radio packets between a gateway and a mobile node~\cite{Sommer2013REALWSN}. A remote procedure call (RPC) is initiated by a radio packet containing the command identifier and optional arguments and is acknowledged by a response packet~\cite{Corke2010}. Our framework provides the flexibility to implement high-level communication protocols for data download, remote configuration or reprogramming on top of basic RPC commands (see Section~\ref{sec:connectivity}). 

\fakeparagraph{Cloud Services (Tier 3)} The third tier in our system consists of a web service located in the cloud. As bandwidth is less restricted between gateway nodes and the cloud services, we employ standard Internet protocols such as HTTP and encode data traffic into JavaScript Object Notation (JSON) objects. At the core of Tier 3 is a REST~\cite{Fielding02} web service which provides access to sensor data, global view of the node download state, and node configuration information. The data ingest service accepts HTTP POST requests containing the sensor data in a JSON object wrapper and stores it using the HDF5 hierarchical storage system. We further implemented a REST API to provide hierarchical access to resources and sensor data, which is also used by our system dashboard for engineers and domain scientists.

\begin{figure}[tbh]
	\includegraphics[width=\textwidth]{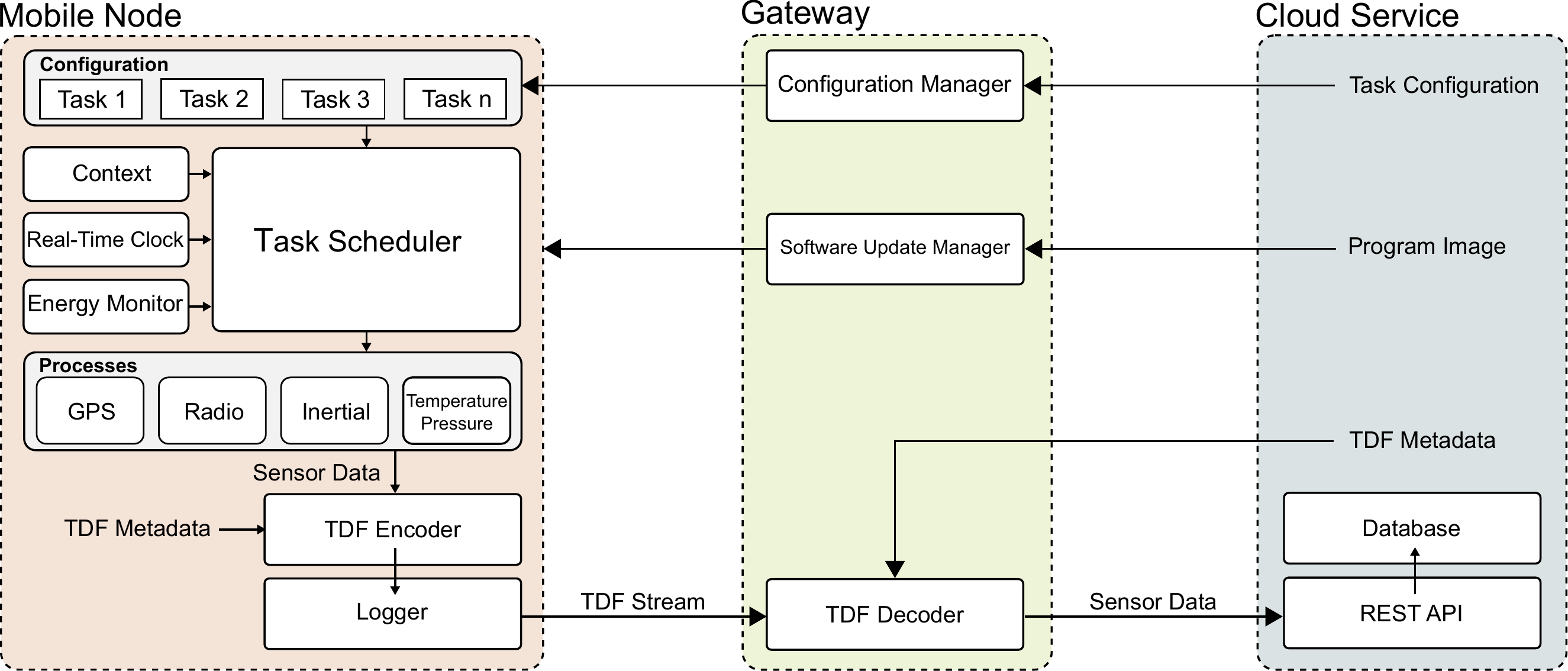}
	\caption{Software architecture of our framework: Sensing tasks can be remotely configured using a set of simple rules that are evaluated at runtime by the energy- and context-aware task scheduler. Sensor readings from different sources are encoded into a TDF stream on the mobile node and decoded on the gateways before the data is forwarded to the cloud service. Gateway nodes can update the task configuration and program image remotely using RPC commands.}
	\label{fig:tdf}
\end{figure}

\section{Energy Constraints}

The main challenge of the software architecture is to integrate devices operating in different system tiers and with different energy budgets. With the exception of the cloud services in Tier 3, individual devices might be duty-cycled and operate according to a schedule that depends on the available energy budget and mobile-node related activity.

\fakeparagraph{Mobile Nodes} The mobile nodes need to operate in a highly efficient manner to meet their strict energy constraints. They must use ultra-low power in sleep mode, optimise sensor sampling for maximum information gain, such as tracking animal location only when the animal is moving, and minimise idle listening in their communication protocols. We designed the application deployed on the mobile nodes to provide unsupervised long-term operation with energy awareness. The battery state of charge is subject to fluctuation, which largely depends on the energy consumed to execute sensing and communication tasks, and on the amount of harvested energy~\cite{Sommer2013ENSSys}. At the core of the mobile node's software is a task scheduler, which manages different sensing tasks based on the available energy and context, as illustrated in Figure~\ref{fig:tdf}.

\begin{figure}
\includegraphics[width=\textwidth]{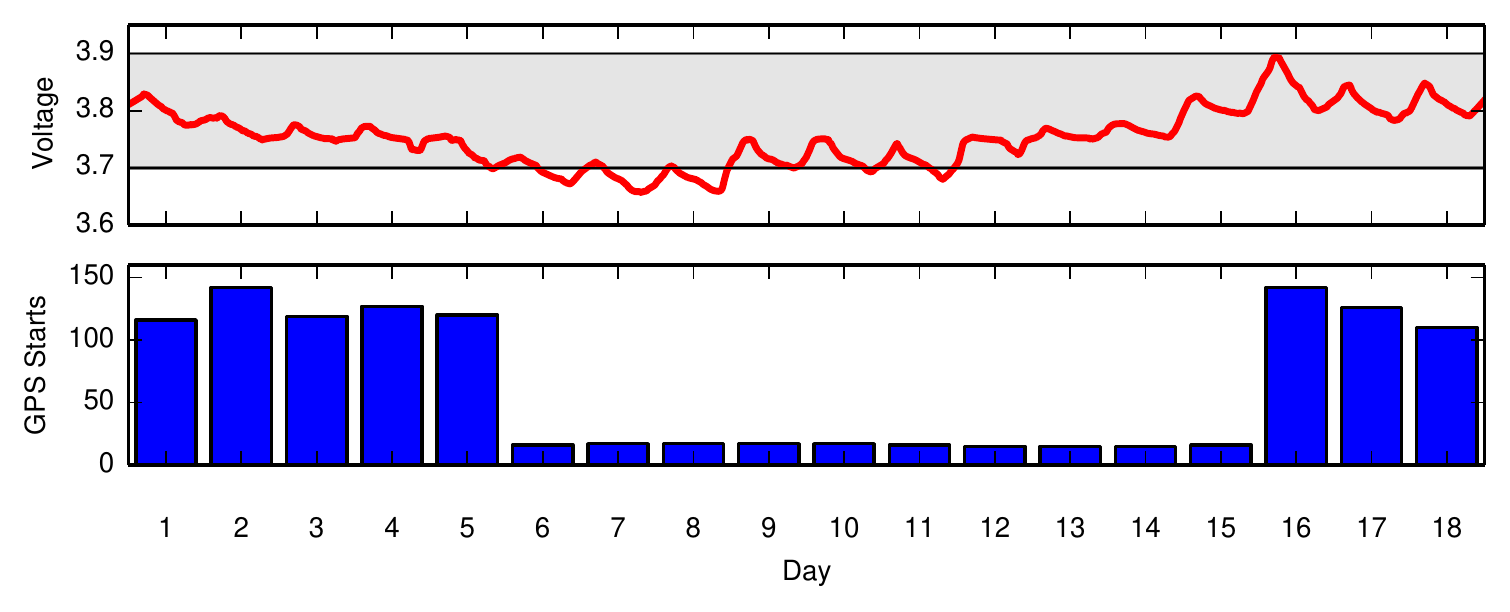}
\caption{Battery voltage measurement and GPS starts as reported by a mobile node attached to a free-living flying fox. The task scheduler decreases the GPS activity as soon as the battery voltage falls below the low voltage threshold of 3.7\,V on Day 6. This allows the battery to recharge and the voltage increases again. GPS sampling returns back to normal sampling as soon as the battery voltage reaches a value of 3.9\,V on Day 16.}
\label{fig:voltage_gps}
\end{figure}

A task configuration defines entry and exit conditions such as time of day, battery voltage and context (\eg animal motion detected) for each sensor process. The task scheduler periodically checks all configurations and starts the corresponding task if its entry conditions match. Similarly, running tasks are checked for the specified exit conditions (\eg number of samples acquired) and stopped if necessary. 
Execution of sensing tasks based on the state of charge allows to gracefully reduce the amount of sensor data collected with decreasing battery charge level. We show an example of this scheduling mechanism for a mobile node deployed on a flying fox in Figure~\ref{fig:voltage_gps}. The scheduler adapts the sampling rate of the GPS receiver based on the measurement of the battery voltage. In case of a low voltage reading, the GPS is sampled less often, \eg only periodic samples are taken instead of motion-triggered continuous tracking. Consequently, more energy harvested from the solar panels can be stored in the battery. Once the battery voltage recovers to an acceptable level, the GPS can be scheduled more often again. Furthermore, multi-modal sensing capabilities (\eg inertial, audio, temperature, pressure) can classify the node context further for optimal scheduling of energy intensive operations, which can result in significant energy savings~\cite{Jurdak2013}.

\fakeparagraph{Gateway Nodes} In general, continuous power supply is not available at suitable gateway locations in animal congregation areas. Therefore, we operate gateways on batteries and employ solar panels for energy harvesting. However, it still might become necessary to operate the embedded PC and cellular network transceiver at a duty cycle, which will result in intermittent connectivity to the cloud services. In order to optimise power usage at each gateway location, we prolong gateway operation when many packets are waiting to be downloaded from mobile nodes, while the gateway can go back to sleep immediately when no mobile nodes are in proximity.

\section{Intermittent Connectivity}
\label{sec:connectivity}

\begin{figure}[tbh]
	\includegraphics[width=\textwidth]{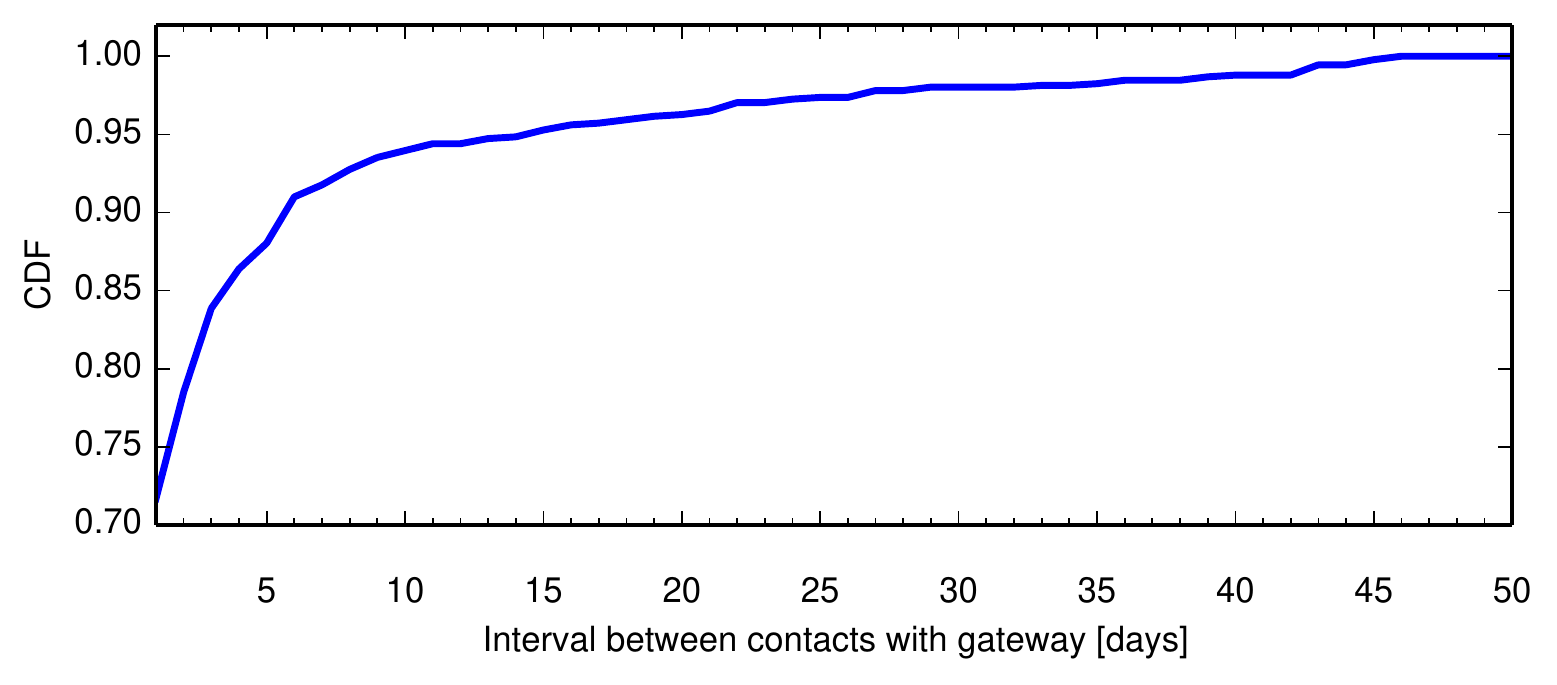}
	\caption{Distribution of the time interval between successive contacts with a wireless gateway for 73 flying foxes (fruit bats) collared with tracking devices.}
	\label{fig:interconnect}
\end{figure}

As mobile nodes spend only short time periods in areas with wireless connectivity, we need to provision for logging sensor data to a temporary storage on mobile nodes and provide delay-tolerant mechanisms to upload the data to the cloud services. For our flying foxes case study, we show the cumulative distribution function (CDF) of the time interval between two subsequent contacts with a gateway node for 73 animals with collars in Figure~\ref{fig:interconnect}. While 70\% of contacts recur daily, the remaining  30\% of inter-contact intervals range between a couple of days to up to six and half weeks, highlighting the need of supporting delay tolerance in the software framework. Furthermore, the framework should support intermittent connectivity not only with the mobile nodes, but also with gateway nodes that may be energy constrained themselves.

\fakeparagraph{Data Storage and Transfer}
We implemented a logging abstraction that allows to access the local storage in a unified way without needing to know the structure of the underlying hardware architecture (\eg Flash chips or microSD cards). The logging abstraction is based on pages, whose size corresponds to the page size of the physical storage medium (\eg 256\,Bytes for common flash chips). We use increasing page numbers to retrieve data from the logger and identify data that has been downloaded already.

The data format used to store sensor readings is an important consideration as it influences the energy efficiency of communicating the data over the radio. For transmission efficiency reasons, we store and transmit data in a binary format, which makes it difficult for humans to interpret directly, but introduces significant energy savings. We use the Tagged Data Format (TDF)~\cite{Corke2010} to encode multiple sensor readings into a byte stream. Each sensor sample is stored with a unique sensor identifier, which defines how the value will be interpreted. All sensor samples are timestamped at a granularity that fits application accuracy needs and storage constraints~\cite{Sommer2013REALWSN}. In line with recent approaches to store and transmit information from Internet of Things devices~\cite{Dawson-Haggerty2010}, we advocate for separation of the actual data from metadata that defines how the data can be interpreted. In our case, the metadata information is stored at the cloud service to which gateways periodically synchronise (see Figure~\ref{fig:tdf}). Mobile nodes, on the other hand, keep a static version of the metadata and backwards compatibility is ensured through creating new data types in the metadata rather than modifying the old ones. 
 
\fakeparagraph{Protocols for Delay-tolerant Network Operation} Organisation of our system into three distinct tiers (mobile nodes, gateways, cloud services) enables us to tailor communication between different tiers to meet the resource constraints imposed by the application scenario. At the mobile node layer, we employ a unicast radio communication protocol based on the Contiki RIME~\cite{Dunkels2004} stack for node discovery and data transfers taking into account the highly asymmetric energy resources available at mobile and gateway nodes. Consequently, mobile nodes employ duty-cycling of the radio transceiver to reduce the power consumption, while gateway nodes can afford to keep their radio enabled for longer periods. Thereby, the mobile node's radio remains powered off most of the time and is only active for a short amount of time to send a packet containing node status information. Unless the gateway requests the mobile node to keep its radio enabled, it will go back to sleep until the next beacon is due.

We implemented a mechanism to download sensor readings from the flash storage using our RPC framework. Data stored in flash pages are sequentially transferred by requesting small chunks that each fits into a single radio packet. In real-world scenarios, we achieve a throughput between 4 and 6 pages downloaded per second. Since data transfers are initiated by the gateway, the application logic in the mobile node is not required to keep track which parts of the buffered sensor data have already been downloaded.
 
\fakeparagraph{Delay-Tolerant Operation with Multiple Gateways} Gateway nodes will forward data downloaded from mobile nodes to a web service, where sensor readings are then stored into the database. Mobile nodes periodically broadcast their maximum available page number, which is overheard by the gateway nodes when in proximity. We use a central web service, which provides the lowest page number that has not been downloaded yet and the range of pages waiting for download. The gateway node will report the successful download of each page to the web service. This ensures that other gateways will not attempt to download the same memory region again in the future.

In addition to data downloaded from mobile nodes, gateway nodes will also periodically report gateway health information to the server such as uptime, battery state of charge, temperature conditions, and storage capacity, which are used to monitor the reliable operation of our network of duty-cycled gateways.

Gateway nodes keep a local copy of the global node state in persistent storage, which allows to operate gateway nodes when no cellular connection to the Internet is available. In this case, sensor data downloaded from mobile nodes is temporarily buffered on the gateway until an Internet connection becomes available and data can be synchronised to the cloud services.

\section{Configuration Management and Remote Debugging}

Long-term tracking operation requires remote inspection and control of the software running on the devices due to lack of physical access after the deployment. For energy efficiency, the system supports reconfigurability at different levels of the software stack: (1) changing parameters such as thresholds, (2) reconfiguration of application-level logic, and (3) replacement of the whole program binary running on the node. Remote instrumentation and debugging techniques have proven to be a crucial tool during the initial prototyping and the actual deployment phases. We leverage our RPC command framework described in Section~\ref{sec:architecture} for interactive memory inspection and to query the application state, such as information about running tasks, battery voltage level, and logger page count.

\fakeparagraph{Remote Configuration} We use a multi-level approach for remote configuration and modification of applications running on inaccessible mobile nodes. We keep a dedicated flash memory area for persistent storage of configuration parameters. Therefore, small changes to the application, such as modifying sensor sampling rates, can be performed without the need for updating the full application code. Furthermore, task configurations can be added, modified or removed using RPC commands. This proved to be a very useful tool when testing and debugging novel software components in the field, as they can be enabled and disabled remotely without the need to update the whole program image.

\fakeparagraph{Wireless Reprogramming} Techniques for remote application updates are known as wireless reprogramming~\cite{Wang06}. In order to update the program code running on the node, we split the binary application image into several smaller chunks that can be transmitted using radio packets. At the node, the image is then re-assembled and written to the program memory. While this approach allows to replace the complete application, a considerable number of radio packets is required to transfer the new image, so it is only used in more significant cases, such as critical bug fixes to low-level components.

\fakeparagraph{Delay-tolerant Configuration Updates} Users might want to change the configuration of one or several nodes to adapt their sensing tasks to a changing context or to address novel research questions. However, the specified nodes might not come into proximity of a gateway for several weeks. Therefore, we use a delay-tolerant approach to update the task configuration and/or the program code running on a node, which does not require manual intervention. Upon receiving a beacon packet from a mobile node, the gateway node requests a description of each node's up-to-date task configuration from the web service and uses RPC calls to compare it to the actual configuration present on the mobile node. If the two configurations differ, the new task configuration is written to the node, and the gateway notifies the cloud service that the node's configuration has been updated. Similarly, we also check for the version number of the program image currently running on a node and update the program binary if necessary.

\section{Discussion and Lessons Learned}

\begin{figure}[tbh]
\centering
\subfigure[]{\frame{\includegraphics[height=4cm]{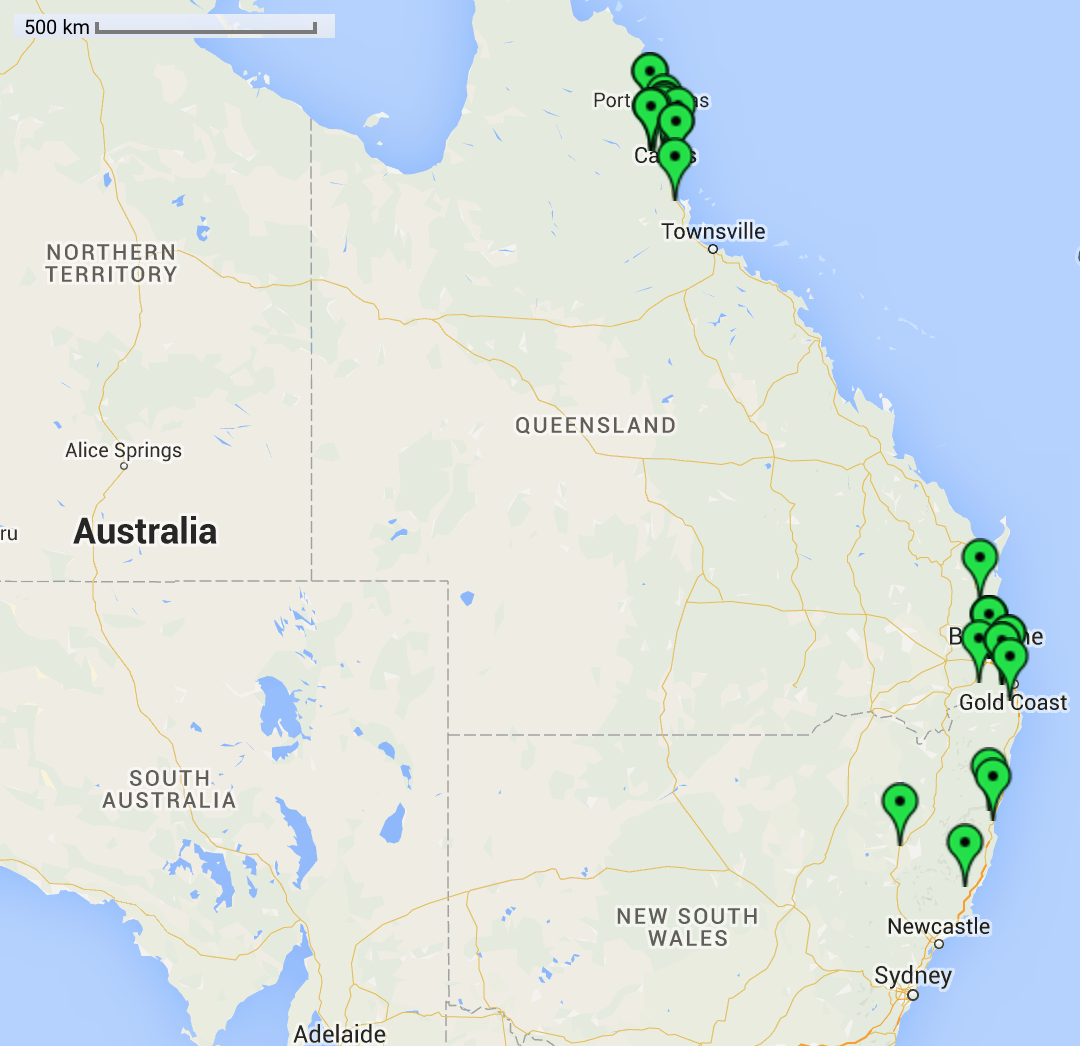}}}
\subfigure[]{\includegraphics[height=4.025cm]{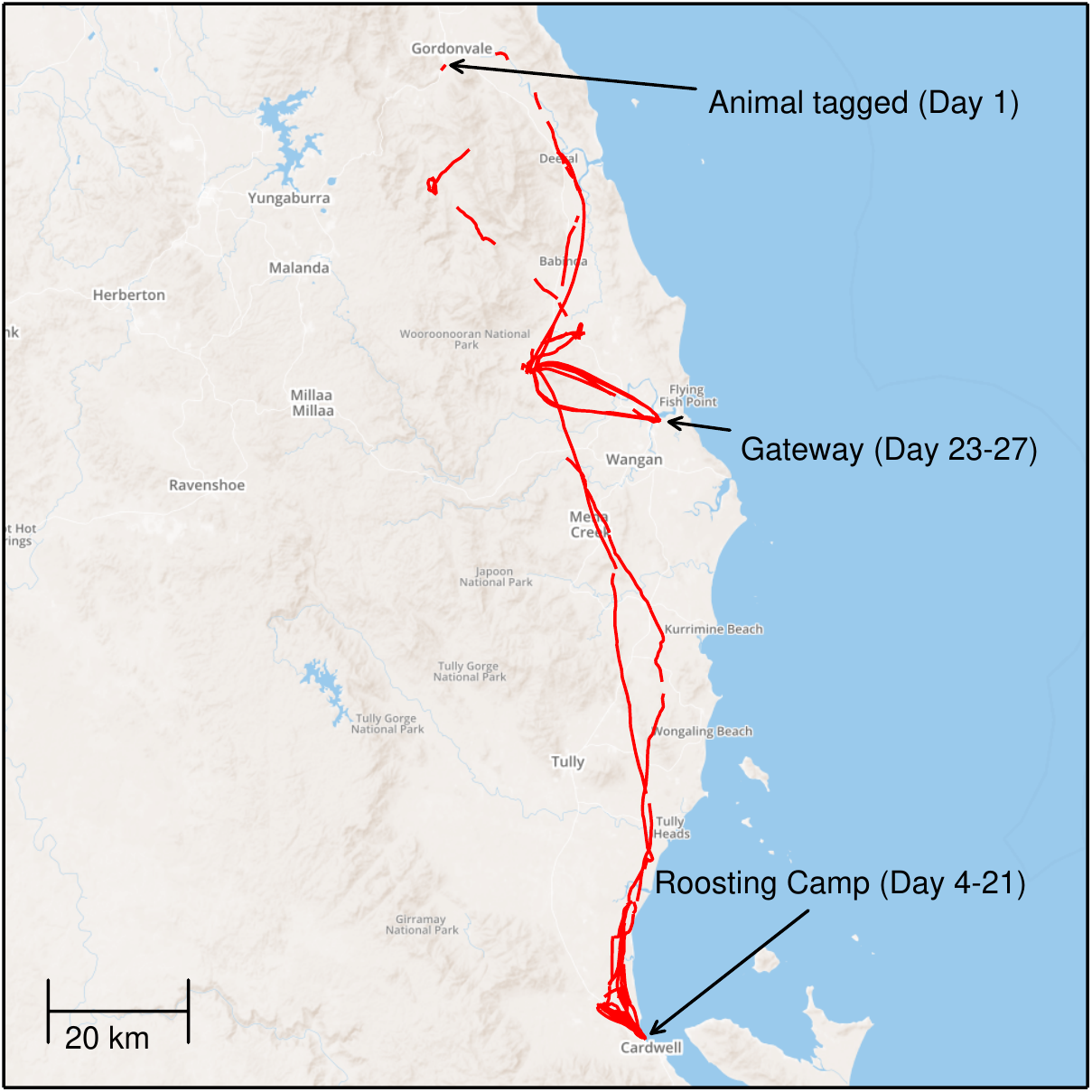}}
\subfigure[]{\includegraphics[width=0.75\textwidth]{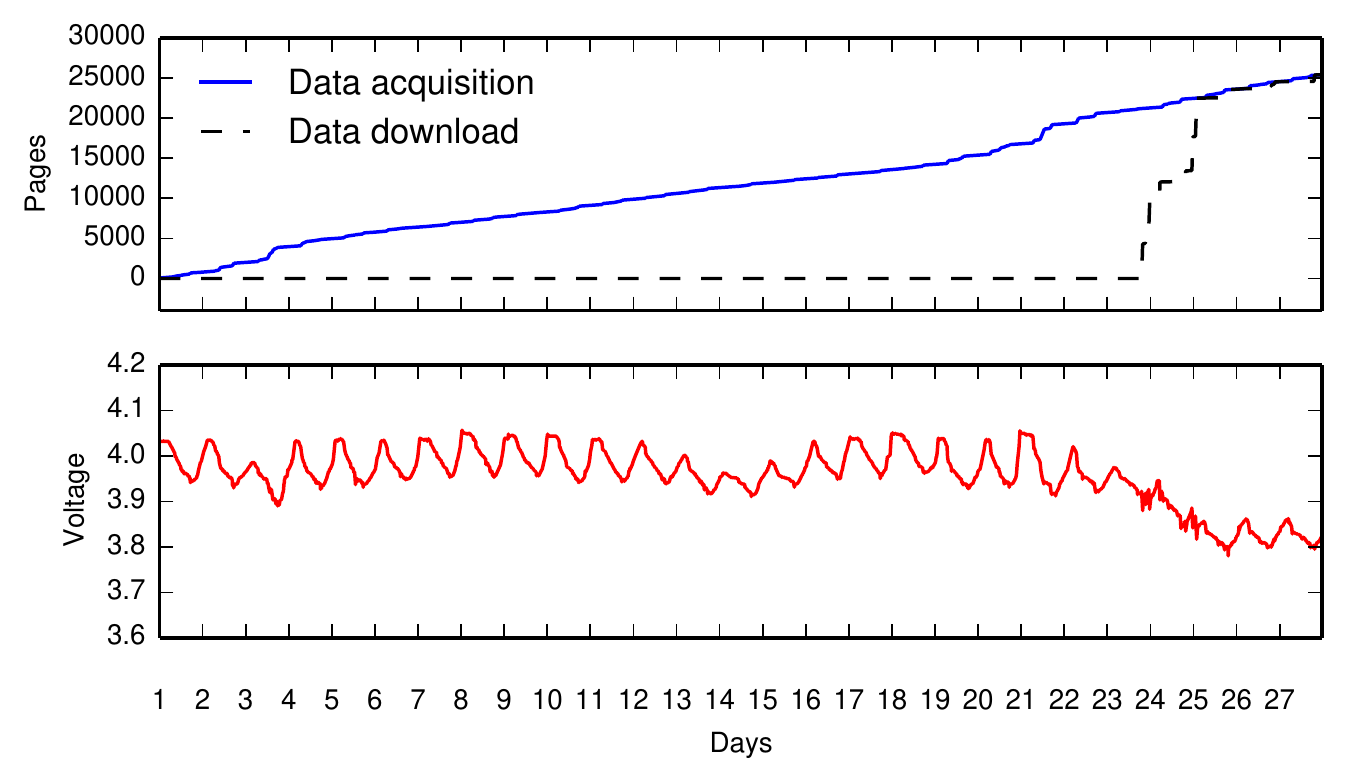}}\\
\caption{(a) network of current gateways across Australia's east coast; (b) movement tracking for an individual collared flying fox based on GPS samples at 1 Hertz; (c) data acquisition and download at the gateway node, and variations in the mobile node's battery voltage during the deployment period.}
\label{fig:example}
\end{figure}

We have successfully implemented the proposed framework in the context of animal tracking and monitoring applications. It has become clear that a multi-tiered architecture is necessary to shift the computational and communication complexity away from inaccessible and resource-constrained devices towards the cloud. We have also established the value of multi-level reconfiguration features for nomadic devices without accessing them physically, ranging from simple parameter updates to full node reprogramming, both in real-time or with delay-tolerance, to maximise versatility as application requirements evolve over time.

Several services and abstractions turned out to be essential during prototyping, testing, and operation of these applications. First, the typed data storage abstraction offered the necessary flexibility with adding different sensing modalities to applications at various sampling rates. Next, the methods that we proposed for autonomous and delay-tolerant operation of mobile nodes proved to be of critical importance for tracking flying foxes, as data had to be buffered locally possibly for weeks at a time until the animal returned to a roosting camp with gateway nodes, as shown in the GPS traces presented in Figure~\ref{fig:example}. It turns out that observed gaps in the trajectory can occur due to spurious mismatches between the sensitivity of the accelerometer-triggered GPS sampling and the accelerometer reading when an animal actually moves, which highlights the importance of remote reconfiguration to optimise these sensitivities over time.

We also learned that while nodes can maintain a healthy battery voltage for a given sensor sampling configuration thanks to solar harvesting, long absences from gateways result in a backlog of data that requires significant energy for download once the node returns (see Figure~\ref{fig:example}(c) after day 23). Sensor sampling frequencies can then be reduced temporarily after this bulk download to allow the node to recover its energy supplies. Adaptiveness of the software provided our domain scientists with an interactive tool to adjust the way the sensors are sampled based on retrospective data analysis or new scientific discoveries, which provides great benefits over deploying several generations of devices during the scientific discovery process.

While animal monitoring has helped motivate and mature our network architecture, we expect it to be useful more broadly, for instance to the traceability of foods products in transit, and more generally to objects that move beyond urban regions.

\section{Acknowledgments}
We would like to thank Ben Mackey, Christopher Crossman and Luke Hovington for their contributions towards the implementation of the software architecture this work is built on. We further thank Adam McKeown, David Westcott, Jiajun Liu and Kun Zhao for their valuable inputs regarding the system architecture for the flying foxes monitoring system. This work has been supported by the CSIRO Sensor and Sensor Networks TCP.

\bibliographystyle{abbrv}
\bibliography{paper}

\end{document}